# Noise-correlation spectrum for a pair of spin qubits in silicon

J. Yoneda[1,2,*], J. S. Rojas-Arias[3], P. Stano[2,4], K. Takeda[2], A. Noiri[2], T. Nakajima[2], D. Loss[2,3,5], and S. Tarucha[2,3,*]

[1] *Tokyo Tech Academy for Super Smart Society, Tokyo Institute of Technology, Tokyo, 152-8552 Japan.*

[2] *RIKEN Center for Emergent Matter Science, RIKEN, Saitama, 351-0198 Japan.*

[3] *RIKEN Center for Quantum Computing, RIKEN, Saitama, 351-0198 Japan.*

[4] *Institute of Physics, Slovak Academy of Sciences, 845 11 Bratislava, Slovakia.*

[5] *Department of Physics, University of Basel, Basel 4056, Switzerland.*

*Correspondence to: yoneda.j.aa@m.titech.ac.jp, tarucha@riken.jp

**Semiconductor qubits are appealing for building quantum processors as they may be densely integrated due to small footprint. However, a high density raises the issue of noise correlated across different qubits, which is of practical concern for scalability and fault tolerance. Here, we analyse and quantify in detail the degree of noise correlation in a pair of neighbouring silicon spin qubits ~100 nm apart. We evaluate all a-priori independent auto- and cross- power spectral densities of noise as a function of frequency. We reveal strong inter-qubit noise correlation with a correlation strength as large as ~0.7 at ~1 Hz (70% of the maximum in-phase correlation), even in the regime where the spin-spin exchange interaction contributes negligibly. We furthermore find that fluctuations of single-spin precession rates are strongly correlated with exchange noise, giving away their electrical origin. Noise cross-correlations have thus enabled us to pinpoint the most influential noise in the present device among compelling mechanisms including nuclear spins. Our work presents a powerful tool set to assess and identify the noise acting on multiple qubits and highlights the importance of long-range electric noise in densely packed silicon spin qubits.**

Precise knowledge about noise is essential for building a quantum computer [1]. The progress in understanding and suppressing noise resulted in qubits with operation errors below 1% in several platforms [2-7]. In these efforts, analysis of temporal correlations or auto-correlations of noise has been instrumental. However, approaches that can disentangle competing remnant error mechanisms with similar or unknown temporal correlations will be desirable. Noise correlations between qubits or qubit Hamiltonian parameters may be promising in this regard. Furthermore, quantum error correction requires the errors to be not only small, but also sufficiently local [8-10]. This crucial requirement for noise correlations across qubits may pose a significant challenge in increasing the number and hence the density of qubits on a quantum processing chip, given the long-range of electric-field fluctuations characteristic of solid-state structures. Quantifying and understanding the noise correlations in densely packed quantum devices is therefore called for [11-13].

In this work, we investigate noise correlations between neighbouring, exchange-coupled, silicon spin qubits separated by about 100 nm [6,7]. We track qubit precession rates in this two-qubit system simultaneously and evaluate the noise correlations between qubits as a function of frequency. Relying on the measured noise correlations, we identify the impact of charge noise that competes with and exceeds that of nuclear spins. Our results include obtaining the cross power spectral densities (cross-PSDs) which – unlike their single-qubit counterparts, the auto power spectral densities (auto-PSDs) – have remained elusive in previous studies [5, 14-16]. Our key observation is that the noise which the two spin qubits see is strongly correlated. We can only interpret this as due to strong influence from correlated fluctuations of electric fields at the qubit positions. We substantiate this interpretation by the observed large noise correlations between single-spin precession rates and the exchange coupling. Furthermore, we find that a simple microscopic device model that attributes all qubit noise to charge noise semi-quantitatively reproduces the observed PSDs.

**Qubit setup and precession rate estimation**

We realize two single-electron spin-qubits, A and B, in an isotopically natural Si/SiGe double quantum dot (see Fig. 1a and Methods). If there were no spin-spin interactions, the two-qubit system would be fully characterized by two precession rates, $\nu_A$ and $\nu_B$. Due to a micromagnet, these two frequencies differ by $|\Delta| \sim 630$ MHz. In addition, once the spin-spin exchange coupling $J$ is non-zero, the precession rate of each qubit depends on the state of the other qubit. The system then displays four – rather than two – precession rates. We denote these four rates as $\nu_Q^\sigma$ with the subscript Q = A/B specifying the precessing qubit and the superscript $\sigma = \uparrow/\downarrow$ the state of the other qubit. In our parameter regime, with $|\Delta| \gg J \sim 1.1$ MHz, $\nu_Q^\sigma$ is essentially given by $\nu_Q \pm \frac{J}{2}$ (+ for $\sigma = \uparrow$ and − for $\sigma = \downarrow$), up to corrections $O(J^2/|\Delta|)$ less than a kHz which we neglect. For example, $\nu_A^\uparrow = \nu_A + \frac{J}{2}$ gives the precession rate of qubit A when qubit B is in the spin-↑ state.

In this configuration, qubit errors are dominated by dephasing, that is fluctuations in the qubit precession rates. For a single qubit, these fluctuations can be experimentally tracked by repeating a Ramsey interference sequence [5]. Once the time-trace of the spin precession rate is obtained in this way, we can analyse single-qubit noise, for example by evaluating its auto-PSD. We need to extend this procedure to access noise correlations, for which the concurrent time-evolution of precession rates of different qubits are required. To this end, we implement four interleaved Ramsey interference sequences with different initial qubit states: ↓↓, ↑↓, ↓↑, and ↑↑. Bayesian estimation on a block of Ramsey data yields a set of four estimates $\nu'^\sigma_Q$ for the four precession rates $\nu_Q^\sigma$ every 60 ms, resulting in time traces such as in Fig. 1b. Details of the interleaved Ramsey measurements and the estimations are in Methods.

We note that the four precession rates, being differences of four energy levels of a two-qubit system, can be parameterized by three numbers. Due to the relations $\nu_Q^{\uparrow/\downarrow} = \nu_Q \pm \frac{J}{2}$ a natural parameter set is $\nu_A$, $\nu_B$, and $J$. As a set of values, the four precession rates are thus informationally overcomplete. However, instead of the true values $v_Q^\sigma$, experimentally we can access only their estimates $v'^\sigma_Q$ which are subject to estimation errors. We distinguish this fact in notation throughout the article, using apostrophes for the latter quantities. We utilize the extra information due to the overcompleteness to correct for these estimation errors, to the extent possible (Supplementary Sec. II gives details on the correction). We also reflect it notationally, putting a tilde over a corrected PSD: For example, $C_{A'B'}$ is the cross-PSD for the precession rates $v'_A$ and $v'_B$ as they were estimated without the correction, while $\tilde{C}_{AB}$ is the corrected spectrum that presents our best statistical inference of a cross-PSD for the experimentally inaccessible true precession rates $v_A$ and $v_B$.

**Noise power spectra**

We first characterize the auto-PSDs of single-qubit noise, to connect to the standard techniques and results. For convenience, we introduce the bare qubit precession rate by $v_Q = (v_Q^\downarrow + v_Q^\uparrow)/2$ (analogous equation holds for quantities with apostrophes). Figure 1c shows the corrected auto-PSDs $\tilde{S}_Q(f)$ together with the 90% confidence intervals (Reference [17] explains how to assign confidence intervals to PSD data; see Methods for an excerpt). Both auto-PSDs display a $1/f$-like decay plus a small Lorentzian part. Single-qubit noise which follows such a $1/f$-like trend has been reported many times and often used as a rationale to attribute it to charge noise. However, this reasoning is undermined by examples of observations assigning $1/f$ fluctuations to nuclei [18] and $1/f^2$ to charge noise [15]. We will show below that one can gain much more insight on the noise nature – whether electric, magnetic, local, global, from device or setup – based on noise correlations.

We next investigate the fluctuations in the exchange coupling $J$. Since the exchange emerges from the Coulomb interaction and the interdot tunnelling, fluctuations of $J$ arise from purely electrical noise. They contribute to two-qubit dephasing, causing correlated qubit phase flips. At the symmetric operation point used in this experiment, $J$ is to the first order decoupled from fluctuations in the double-dot detuning [19]. However, it is still sensitive to fluctuations in the tunnel coupling due to changes in the double dot potential landscape. Noting that $\nu_Q^\uparrow - \nu_Q^\downarrow = J$ holds for both qubits, we define the estimator $J' \equiv (\nu'^\uparrow_A - \nu'^\downarrow_A + \nu'^\uparrow_B - \nu'^\downarrow_B)/2$. Figure 1c shows the corrected auto-PSD $\tilde{S}_J(f)$. As expected, it follows a $1/f$ dependence at low frequencies before it hits the noise floor at $\sim 30$ kHz$^2$/Hz. We note that the confidence interval assures us that the observed trend is not an artefact, even for $S_{J'}(f)$ only barely larger than the noise due to the estimation errors and thus our detection procedure itself (being $\sim 300$ kHz$^2$/Hz, as best seen from the frequency plot of $S_{Z'}$ given in Supplementary Fig. S3). We conclude that $S_J(f)$ is more than two orders of magnitude smaller than the single-qubit auto-PSDs, and the qubit errors caused by fluctuations of $J$ will be negligible.

**Quantifying noise correlation**

We now turn to the correlation between single-qubit precession rates as another source of correlated qubit errors. The correlation will be best quantified by $c_{AB}(f) \equiv C_{AB}(f)/\sqrt{S_A(f)S_B(f)}$: a cross-PSD between $\nu_A$ and $\nu_B$, denoted by $C_{AB}(f)$, normalized to the geometric mean of the related auto-PSDs. Unlike auto-PSDs, cross-PSDs are complex even for real-valued variables. The phase $\text{Arg}[c_{AB}(f)]$ signifies the nature of the correlation at the given frequency: the angle 0 means fully in-phase correlation and the angle $\pi$ means fully out-of-phase correlation. Intermediate angles can be interpreted as a time lag of $\nu_A$ with respect to $\nu_B$ measured in units of $1/f$. The amplitude of $c_{AB}(f)$ gives the correlation strength ranging between 0 (no correlation) and 1 (perfect correlation).

Figures 2a and b show the amplitude and phase of $c_{A'B'}(f)$, the normalized cross-PSD for $\nu_A'$ and $\nu_B'$. The correlation is finite at any reasonable confidence level for most frequencies. We can identify two regimes with a crossover taking place at 40-80 mHz. At frequencies below 40 mHz, the qubit precession frequencies fluctuate out-of-phase, $\text{Arg}[c_{A'B'}] \approx \pi$, and they become in-phase above 80 mHz, $\text{Arg}[c_{A'B'}] \approx 0$. Around 1 Hz, the correlation strength is as large as ~0.7 (reaching a maximum of 0.70 ± 0.06 at 1.32 Hz). In sum, the data clearly show that the precession-rate fluctuations, and hence the qubit errors, are strongly correlated.

**In-phase and out-of-phase fluctuations**

Clear in-phase and out-of-phase correlations observed in Fig. 2b suggest looking into the sum and difference of the rates: $\Sigma \equiv \nu_A + \nu_B$ and $\Delta \equiv \nu_A - \nu_B$. We plot their corrected auto-PSDs, $\tilde{S}_\Sigma(f)$ and $\tilde{S}_\Delta(f)$ in Fig. 3. Note that when $\nu_A$ and $\nu_B$ are uncorrelated, $S_\Sigma$ equals $S_\Delta$ (up to statistical fluctuations), whereas the converse is not necessarily true. We find that the magnitude relation between $\tilde{S}_\Sigma$ and $\tilde{S}_\Delta$ above and below the crossover at 40-80 mHz is in line with the conclusions from Fig. 2: $\tilde{S}_\Sigma$ ($\tilde{S}_\Delta$) is larger when the noise is positively (negatively) correlated. When we translate these PSDs to the dephasing times for even (odd) parity Bell states [16] for an integration time of 100 sec by integrating the best fit results for $\tilde{S}_\Sigma$ ($\tilde{S}_\Delta$), we obtain $T_2^* = 0.87$ μs (1.3 μs), with a parity dependence as large as 50%. The strong positive correlation at around 1 Hz can be understood as due to the Lorentzian component [20] in $\tilde{S}_\Sigma$: $0.5\, b^2 \tau_0 / (1 + (2\pi f \tau_0)^2)$ with $b = 282$ kHz and $\tau_0 = 0.162$ sec (plotted along with the difference $\tilde{S}_\Sigma - \tilde{S}_\Delta$ in the inset of Fig. 3). The only plausible explanation we have for this Lorentzian part is charge noise – caused by a two-level charge impurity with switching time $\tau_0$, shifting precession rates of both qubits equally by $b$.

Other than Lorentzian parts, $\tilde{S}_\Sigma$ and $\tilde{S}_\Delta$ are well fitted by a power-law $1/f^\gamma$ with $\gamma \approx$ 1.2-1.3. We found similar auto-PSDs in other quantum-dot devices fabricated with different gate structures on the same isotopically natural Si/SiGe wafer (data not shown). In addition, the dephasing times $T_2^*$ of qubits A and B are ~ 1.1 μs and 1.2 μs (1.3 μs and 1.4 μs) for an integration time of 100 seconds, which are obtained directly from Ramsey measurements (by integrating auto-PSDs). This independence of the gate layout and the values of $T_2^*$, both based on the analysis of single-qubit noise (auto-PSDs), make an argument in favour of isotopically naturally abundant $^{29}$Si nuclear spins as a relevant source of the local noise. However, noise correlations tell us a different story. First, our theoretical analysis (Supplementary Section IV) concludes that they cannot be accounted for by thermally diffusive nuclear spins. On the contrary, power-law behaviour of $\tilde{S}_\Sigma$ and $\tilde{S}_\Delta$ is naturally explained by spatially correlated charge noise that shifts the qubits in space in-phase and out-of-phase in the presence of the magnetic field gradient (Supplementary Section III). In this scenario, the cross-over at 40-80 mHz is the change in the dominance from the in-phase to out-of-phase fluctuations. Therefore, the character of correlations suggests that they originate in charge noise. Indeed, by analysing noise correlation spectra, we can furthermore show that charge noise is a major contributor to the qubit noise in our device as we discuss next.

**Electrical fluctuation and qubit noise spectra**

What is arguably the most telling about the significant contribution of charge noise to qubit dephasing is the correlation of the noise in bare qubit precession rates and the qubit-qubit exchange. This correlation is quantified by cross spectra $c_{Q'J'}$. As shown in Fig. 4a, the normalized correlation strength $|c_{A'J'}|$ is as large as ~0.8 at low frequencies (see also Supplementary Fig. S4 confirming that $\tilde{C}_{AJ}$ is comparable in size with the geometric mean of $\tilde{S}_A$ and $\tilde{S}_J$). While, apart from the charge noise, $\nu_A$ is also subject to nuclear-spin noise, the fluctuations in $J$ are unambiguously electrical in origin. Therefore, the observation of strong correlation proves that noise in $\nu_A$ is also dominated by charge noise.

This conclusion is reinforced by our finding that $c_{A'J'}$ can be predicted from other spectra assuming they all stem from fluctuations of local electric fields seen by the qubits. We reflect this assumption by a simple model which works as follows. The basic quantities of the model are the electric-field fluctuations at individual qubit locations, $\delta E_A$ and $\delta E_B$. They are described by three spectra, $S_{E_A}$, $S_{E_B}$, and $C_{E_A E_B}$. The six observed spectra, $\tilde{S}_A$, $\tilde{S}_B$, $\tilde{S}_J$, $\tilde{C}_{AB}$, $\tilde{C}_{AJ}$, and $\tilde{C}_{BJ}$, can be expressed in terms of the electric-field noise using susceptibilities of $\nu_A$, $\nu_B$, and $J$, which we can estimate from simulation and independent analysis (see Supplementary Section III and Fig. S6 for details of the modelling and the determination of susceptibilities). Having the theoretical relations between the two sets of spectra, we obtain the model spectra $S_{E_A}$, $S_{E_B}$, and $C_{E_A E_B}$ from the observed spectra $\tilde{S}_A$, $\tilde{S}_B$, and $\tilde{C}_{AB}$ (see Supplementary Fig. S7 for the resulting model spectra). After that, we predict $S_J^E$, $C_{AJ}^E$, and $C_{BJ}^E$, the auto-PSDs of the exchange and the cross-PSDs between bare qubit precession rates and the exchange. We find that they reproduce the measured PSDs well, see Supplementary Fig. S4 (note that these results are essentially free from fitting parameters, see Supplementary Sec III). Plotting their normalized versions in Fig. 4, we see an excellent agreement with the experimental observation $c_{A'J'}$. We point out that since $J$ is electric in origin, other sources of noise, if present in the device, will only reduce the correlation strength, $|c_{Q'J'}|$. This is the case for qubit B: applying the same procedure for it, $c_{BJ}^{E}$ overestimates the measured amplitude $|c_{B'J'}|$, suggesting that $\nu_B$ is subject to additional noise. We believe that it comes from nuclear spins, making the prediction for $c_{B'J'}$ correct only semi-quantitatively.

We would like to stress the following aspect. In the above, we have converted the correlated noise of three qubit-related variables ($\nu_A$, $\nu_B$, and $J$) to correlated noise of two local electric fields ($\delta E_A$ and $\delta E_B$). Rather than the number of parameters, the crucial difference is that the latter set is detached from the details of quantum dots and qubits. Spin-qubit electrical noise is usually reported in terms of effective gate-voltage or quantum-dot detuning [20]. These quantities are particular to a given qubit implementation and hard to translate to a different one. In contrast, knowing local electric field noise is much more generic. It allows one to, for example, estimate coherence times for different qubit types (single-spin, singlet-triplet, hole, or charge) [21, 22] or assess device quality irrespective of qubit particulars. Importantly, it also provides hints on the location of noise sources, especially if a large correlation or anti-correlation is seen. Taken together, we believe that description in terms of electric fields provides clear advantages, especially for devices with multi-qubit arrays.

To conclude, we have analysed the degree of noise correlations in a neighbouring spin-qubit pair. An important technical part of the analysis was the application of Bayesian estimation of auto- and cross-PSDs recently developed in Ref. [17]. It allowed us to quantify the statistical relevance of the observed correlations and confirm beyond doubt that the two qubits see a noise which is strongly correlated. Importantly, this correlation does not arise through the fluctuations in the spin-spin exchange: the qubits see correlated noise even if they do not interact. The second essential finding is that all the observed noise correlations and large portions of the noise local components are of electrical origin (it enters the spin-qubit precession rates through the micromagnet-induced magnetic-field gradients in our work, but similar mechanisms exist for other electrically tunable spin qubits, such as spin-orbit interactions or the Stark effects). These findings have implications on the possibilities of the noise mitigation. The important open question is how far the correlation will extend in a qubit array. (We would like to note that this problem of spatial correlation of noise is distinct from the technical challenge of mitigating qubit crosstalk [23, 24].) Our observations from qubits separated by ~100 nm suggest that the noise correlation due to electrical noise will not likely decay at least on the length scale of tens of nanometers. The scaling with distance will be critical concerning quantum-error-correction protocols and fault-tolerant architectures [8, 25] as well as the identification of the physical source of this electrical noise. We believe that the methods of including cross-PSDs in noise analysis that we pioneered here experimentally building upon a recent theoretical development [17] provide crucial tools for answering such questions.

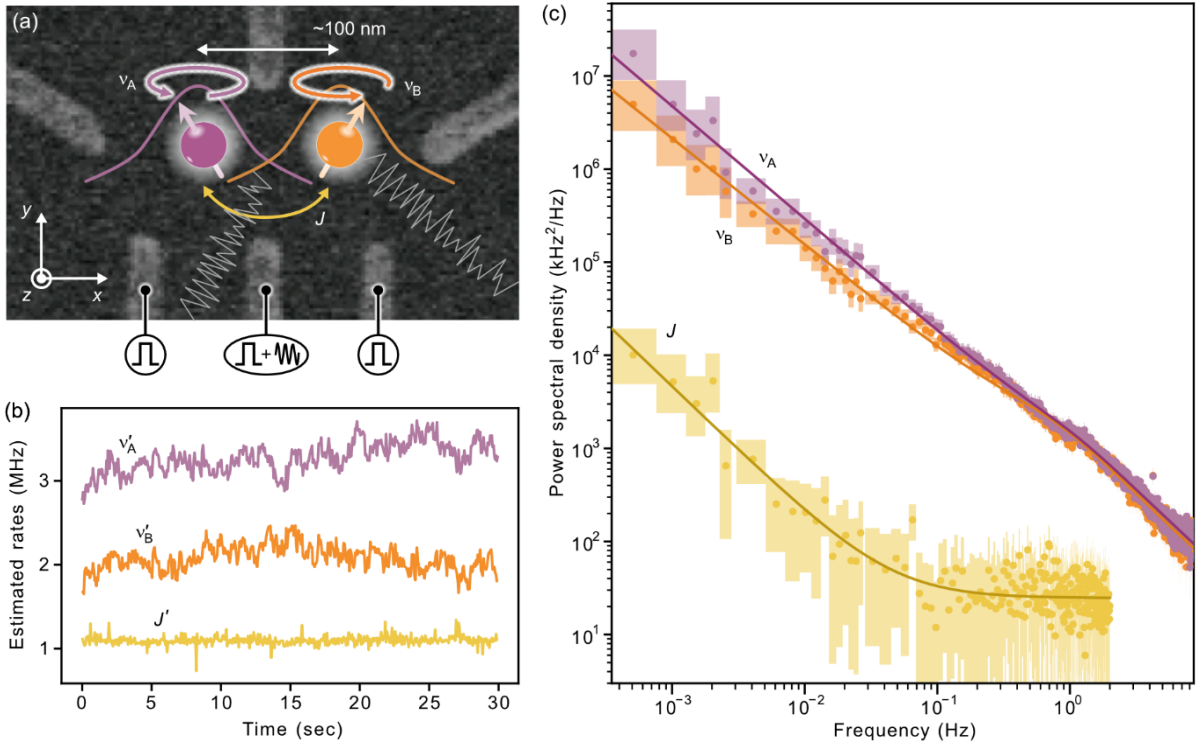

**Fig. 1. Qubit device and power spectral densities. (a)** Schematic representation of the two-qubit device with a scanning electron micrograph of a nominally identical device. Qubits are ~100 nm apart, and are coupled by the spin-spin exchange $J$. They precess in the field of the external magnet and the micromagnet. The precession rates are perturbed by noise, illustrated by the grey zigzag lines. **(b)** An example of time traces of qubit precession rates and exchange, all simultaneously measured as explained in the main text. Traces are offset for clarity. **(c)** Auto-PSDs $\tilde{S}_A$ (purple), $\tilde{S}_B$ (orange), and $\tilde{S}_J$ (yellow) corrected for rate estimation errors. We use a Bayes procedure to get posterior probability distributions [17] for auto-PSD at a given frequency. Dots give the means and shaded regions the 90% confidence intervals associated to these posteriors. $\tilde{S}_J$ is overwhelmed by the noise floor at high frequencies and not plotted above 2 Hz to avoid visual overlap with $\tilde{S}_Q$. Purple and orange curves show fit results to a power law plus a Lorentzian: $a(f/1\text{ Hz})^{-\gamma} + 0.5\, b^2\tau_0/(1+(2\pi f\tau_0)^2)$. The reference frequency of 1 Hz is introduced to simplify the units of $a$. The best-fit values for qubit A (B) are $a = 1100$ kHz$^2$/Hz (800 kHz$^2$/Hz), $\gamma = 1.21$ (1.14), $b = 105$ kHz (129 kHz) and $\tau_0 = 0.140$ sec (0.175 sec). The yellow curve is a fit to $a(f/1\text{ Hz})^{-\gamma} + g$, with best-fit values $a = 0.36$ kHz$^2$/Hz, $\gamma = 1.37$, and $g = 25$ kHz$^2$/Hz.

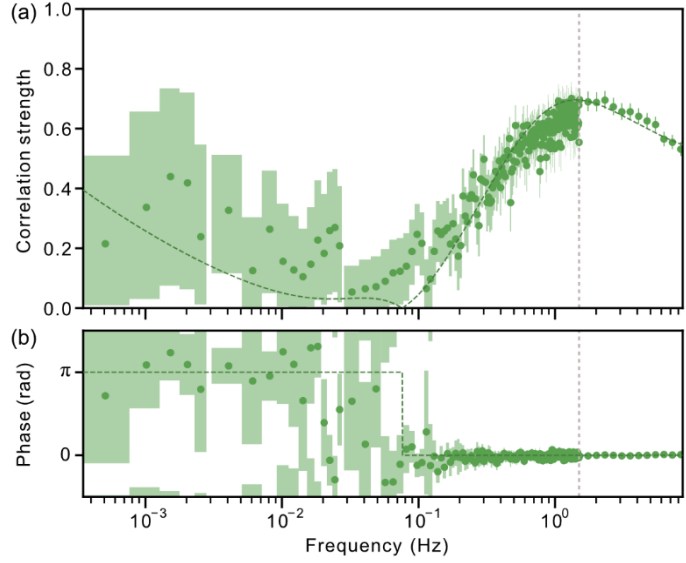

**Fig. 2. Normalized cross-PSD between bare precession rates. (a)** Correlation strength and **(b)** phase. Dots show the posteriors means and shaded regions the 90% confidence intervals. The vertical dashed line at 1.5 Hz separates regions where different methods are used for calculation. To the left of the line, we plot $c_{A'B'}$, the un-corrected cross-PSD between $\nu'_A$ and $\nu'_B$. To the right to the line, we plot the ratio $\tilde{C}_{AB}/\sqrt{\tilde{S}_A \tilde{S}_B}$, with the correction for the rate estimation errors applied individually for the three PSDs involved (However, note that $C_{A'B'} = \tilde{C}_{AB}$, meaning it is unaffected by the correction, see Supplementary Section II). We find that this way of presentation is best in suppressing the artifacts of the rate estimation errors which otherwise lead to an underestimation of the correlation strength (demonstrated in Supplementary Fig. S5) due to an overestimation of the terms in the denominator. This conclusion is supported by comparing the data to the dashed curves, which give the results according to the simplified scenario allowing only either completely in-phase or out-of-phase cross-correlation (see "Calculation of normalized spectra" in Methods). In addition, to improve the plot readability in this frequency region, we merge probability distributions at neighbouring frequencies, following the procedure given in Ref. [17].

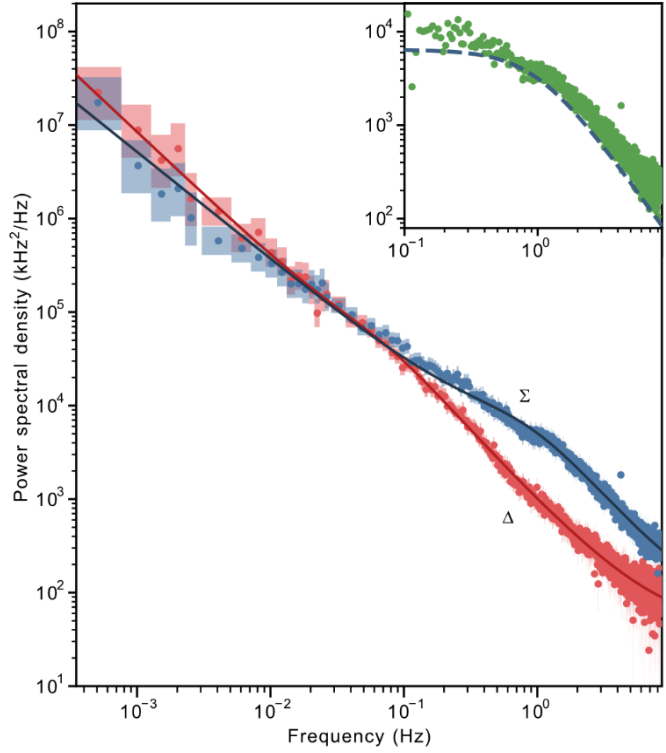

**Fig. 3. Corrected auto-PSDs of the sum (Σ) and the difference (Δ) of the bare precession rates.** Dots give the posterior means and shaded regions the 90% confidence intervals. Solid curves show the fit to $a(f/1\text{ Hz})^{-\gamma} + 0.5\, b^2\tau_0/(1 + (2\pi f\tau_0)^2) + g$ yielding $a = 1860\text{ kHz}^2/\text{Hz}$ (785 kHz$^2$/Hz), $\gamma = 1.15$ (1.34), $b = 282$ kHz (182 kHz), $\tau_0 = 0.162$ sec (2.2 sec) and $g = 43$ kHz$^2$/Hz for $\tilde{S}_\Sigma$ ($\tilde{S}_\Delta$). When we fit $\tilde{S}_\Sigma$, $g$ is fixed to the best-fit value obtained from the fit to $\tilde{S}_\Delta$. The inset shows the difference in the probability distribution mean between $\tilde{S}_\Sigma$ and $\tilde{S}_\Delta$ (dots), along with the Lorentzian component of the best fit for $\tilde{S}_\Sigma$ (dashed line).

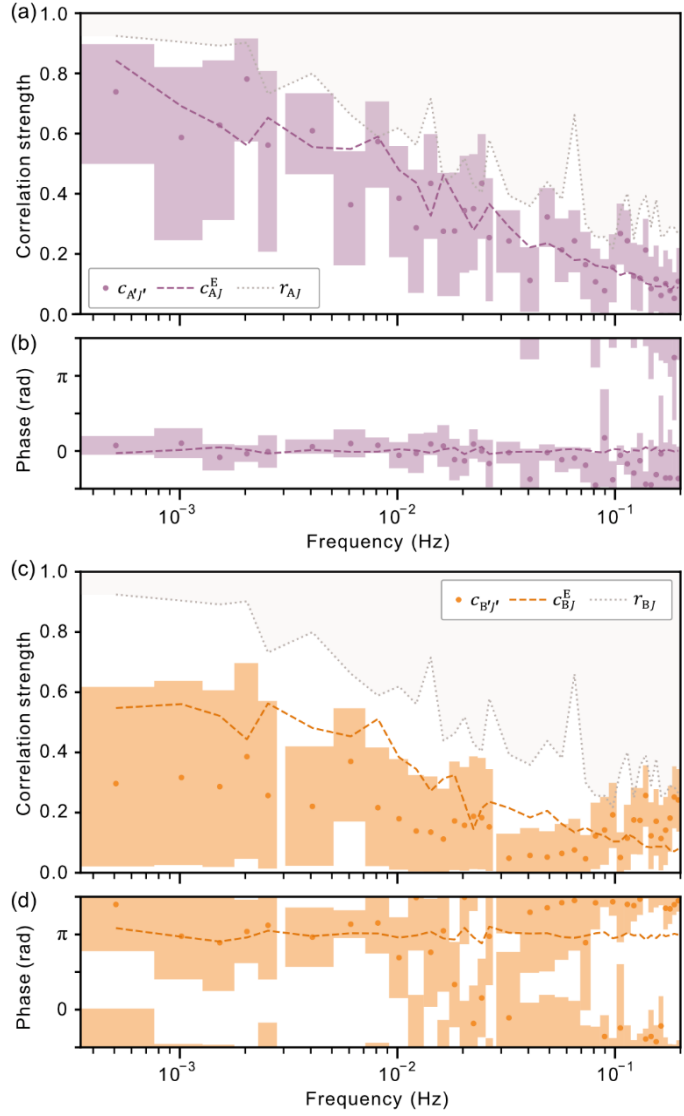

**Fig. 4. Normalized cross-PSD between a bare precession rate and the exchange. (a)** Correlation strength and **(b)** phase of the normalized cross-PSD, $c_{A'J'}(f)$, plotted with the result from electric field model, $c_{AJ}^{E}(f) \equiv C_{AJ}^{E}(f)/\sqrt{S_{A'}S_{J'}}$. The expression $r_{AJ}(f) \equiv \sqrt{\tilde{S}_A \tilde{S}_J}/\sqrt{S_{A'}S_{J'}}$ quantifies the influence of rate-estimation errors in the denominator. Dots give the posterior means and shaded regions the 90% confidence intervals. No correction for the rate estimation errors is done for $c_{A'J'}(f)$. **(c), (d)** Corresponding data for $c_{B'J'}(f)$. $c_{BJ}^{E}$ is slightly larger than $c_{B'J'}$ due presumably to uncorrelated nuclear spin noise (not included in the model).

## Methods

**Cross correlation and related terminology**

We use the word 'cross-correlation' to mean the property defined on a pair of signals, $\alpha$ and $\beta$, related to their correlation, or "degree of similarity", quantified by the cross-PSD $C_{\alpha\beta}$. The latter is defined by

$$C_{\alpha\beta}(f) \equiv \int_{-\infty}^{\infty} d\tau\, e^{2\pi i f \tau} \langle \alpha(t)\beta(t+\tau) \rangle,$$

with $\langle \cdots \rangle$ denoting the statistical average (the auto-PSD is defined by setting $\alpha = \beta$). This quantity is complex, so it has a magnitude and a phase. The two signals are "correlated" if the magnitude is non-zero and "uncorrelated" if it is zero (more precisely, below a chosen threshold). If the two signals are from different locations in space, one sometimes talks about "spatial correlation". When the correlation is finite, the phase gives further information on the correlation nature: the phase close to zero is referred to as "in-phase" or "positively correlated" signals, while phase close to $\pi$ is referred to as "out-of-phase", "negatively correlated" or "anti-correlated" signals. The possible confusion with these technical terms arises with some existing literature which uses "correlated" to what we call here "positively correlated" without giving the adverb.

**Experimental setup**

The qubit device is fabricated on an isotopically natural Si/SiGe heterostructure wafer and is measured in the same setup as Ref. [26]. The external magnetic field of 0.51 T translates into the qubits with frequencies (called precession rates in the main text) of ~16.3 GHz, which are further split by ~620 MHz due to the magnetic field gradient created by a micromagnet. The qubits can be manipulated via the exchange interaction and electric-dipole-spin-resonance. We tune the exchange coupling $J$ to 1.1 MHz and work at a so-called symmetric operation point where $J$ is least sensitive to electrical noise. Single qubit $\pi/2$ rotation times are 65.5 ns (103.5 ns). The microwave is synthesized from two Keysight E8267D signal generators with the IQ modulation signal sent from a 4-channel AWG, Tabor Electronics WX2184. From the thermal broadening of charge transition lines, we estimate the electron temperature to be ~50 mK. The qubit readout relies on rf-reflectometry charge sensing, detecting spin-dependent tunnelling to reservoirs [27]. The sensing signal is sampled by an AlazarTech digitizer ATS9440 at 10 MSPS, filtered at 1 MHz using a second order Butterworth digital filter and decimated at 2 MSPS for post processing. We record the difference between the maximum and the minimum readings within the 45 μs readout window.

Some qubit noise spectra show a spike at 4.2 Hz (see also Supplementary Fig. S9 and Section V). This feature is not an artefact, judging from the corresponding confidence intervals. We attribute the peak to mechanical vibrations due to the pulse-tube cooler, Cryomech PT410 [28], installed in the dry dilution refrigerator, Oxford Instruments Triton200. It corresponds to the third-order harmonics of its 1.4 Hz operation frequency. No other harmonic frequency is visible above the noise level due to other noise sources. Inside the solenoid magnet, vibration noise should be converted to fairly spatially-uniform magnetic noise. Indeed, the spike is absent from $\tilde{S}_\Delta(f)$ and observed exclusively in $\tilde{S}_\Sigma(f)$ in Fig. 3. Note that the triboelectric effect in the cables as previously discussed in the literature [29] would in contrast contribute to both $\Sigma$ and $\Delta$, and thus is ruled out by the measured noise correlations.

**Qubit frequencies**

In a system of two exchange-coupled qubits, A and B, the precession rates can be expressed as

$$\nu_Q^\downarrow = \frac{\Sigma \pm \sqrt{\Delta^2 + J^2} - J}{2} \approx \nu_Q - \frac{J}{2} \pm \frac{J^2}{4|\Delta|}$$

$$\nu_Q^\uparrow = \frac{\Sigma \pm \sqrt{\Delta^2 + J^2} + J}{2} \approx \nu_Q + \frac{J}{2} \pm \frac{J^2}{4|\Delta|}$$

where different signs correspond to different qubits, $\Sigma \equiv \nu_A + \nu_B$, and $\Delta \equiv \nu_A - \nu_B$. Note that in our device $J/(4|\Delta|) < 0.0005$ so that $\nu_Q^{\downarrow(\uparrow)}$ is well approximated by $\nu_Q \pm \frac{J}{2}$.

**Interleaved Ramsey sequence**

In order to track simultaneously several qubit-precession-rates, we interleaved four different sequences of repeated Ramsey interferences. Each of the four sequences corresponds to one of the four rates, $\nu_Q^\sigma$, where qubits are prepared in one of the four eigenstates by turning on and off 10 μs-long adiabatic inversion microwave pulses. Each sequence comprises four repetitions of 100 pulse cycles with different values of the evolution time (linearly changed from 0.02 to 2 μs). Each pulse cycle lasts 150 μs, most of the time spent in the measurement, initialization and DC-bias compensation, and yields one spin-readout signal for each qubit. To ensure simultaneity, pulse cycles from four different sequences are interleaved, such that the qubit initial states are switched before the evolution time is changed. Combined with the Bayesian model (next section), we achieve an estimated average power of the rate estimation errors of ~300 kHz$^2$/Hz (Supplementary Fig. S3).

**Bayesian model for qubit precession rate estimation**

We construct the Ramsey oscillation model which relates the readout outcome with the qubit detuning in relation to the control microwave tone. The following practices are employed to reduce the number of samples required to obtain a frequency estimator. First, in the Ramsey oscillation model, we take into account the empirically observed linear dependence of the oscillation parameters (for example, phase) on the qubit detuning (Supplementary Fig. S2). Second, we set the Bayes prior to the normal distribution with a standard deviation of 100 kHz with its mean at the value estimated from the 16 preceding sequences. Third, we use the sensor signal histograms (estimated similarly to Ref. [26]) as inputs to the Bayesian estimation based on the Ramsey oscillation model (see Supplementary Figure S1 and Section I for details).

**Calculation of unnormalized spectra**

We follow the procedures detailed in Ref. [17], which yield statistical uncertainty measures based on Bayesian probability theory, when we calculate the unnormalized spectra (both auto- and cross-PSDs). We start with 8 blocks of data each containing 32752 sets of precession rates, which we group into $M = 8$, 32 or 128 batches of $N = 32752$, 8188 or 2047 points for a frequency range below 2.7 mHz, between 2.7 mHz and 27 mHz, or above 27 mHz, respectively, unless otherwise noted. We note that we take into account the errors in the precession rate estimation. We achieve noise floors of the corrected spectra on the order of ~100 kHz$^2$/Hz (they depend on the variables). The details of this subtraction process can be found in Supplementary Section II. We plot the mean and 90% confidence intervals determined from the Bayes posterior probability distributions.

**Calculation of normalized spectra**

Due to high computational costs, we cannot correct for the rate-estimation errors the amplitudes of normalized cross-PSDs, that is the correlation strengths plotted in Figs. 2 and 4. At low frequencies, the lack of correction has no consequences. It starts to matter once at least one of the PSDs becomes comparable or lower than the noise due to rate-estimation errors. Once this happens, the (uncorrected) correlation strength tends to be underestimated since the auto-PSD(s) in the normalizing denominator is overestimated. To mitigate this problem in Fig. 2, we evaluate the correlation strength using $|\tilde{C}_{AB}|/\sqrt{\tilde{S}_A \tilde{S}_B}$ instead of using the uncorrected $c_{AB}$ above 1.5 Hz. The consistency of plotting the results in this way can be tested. Namely, if $c_{AB}$ is strictly real, it can be simplified to $\frac{1}{4}(S_\Sigma - S_\Delta)/\sqrt{S_A S_B}$. We plot the amplitude and phase (being 0 or $\pi$) resulting from this formula evaluated using $S_\Sigma$, $S_\Delta$, $S_A$, and $S_B$ as fitted in Figs. 1(c) and 3. A good agreement between the two ways of evaluation reassures us that the way of plotting the results above 1.5 Hz in Fig. 3 is sound.

**Acknowledgements** We are grateful to Á. Gutiérrez-Rubio for fruitful discussions. Part of this work was financially supported by MEXT Quantum Leap Flagship Program (MEXT Q-LEAP) grant no. JPMXS0118069228, JST PRESTO grant no. JPMJPR21BA, JST Moonshot R&D grant no. JPMJMS2065, JST CREST grant no. JPMJCR1675, JSPS KAKENHI grant no. JP21K14485, Swiss National Science Foundation (SNSF) and NCCR SPIN, The Precise Measurement Technology Promotion Foundation, Suematsu Fund, Advanced Technology Institute Research Grants.